\documentclass{pasj00}
\draft

\begin{document}
\SetRunningHead{Yonehara, Umemura, \& Susa}{Quasar Mesolensing}
\Received{2003/08/22}
\Accepted{2003/10/10}

\title{Quasar Mesolensing - Direct Probe to Substructures around Galaxies -}

\author{Atsunori \textsc{Yonehara}%
  \thanks{Research Fellow of Japan Society for Promotion of Science}}
\affil{Center for Computational Physics, University of Tsukuba, \\
Tennoudai 1-1-1, Tsukuba, Ibaraki, 305-8577, JAPAN}
\email{yonehara@rccp.tsukuba.ac.jp}

\author{Masayuki \textsc{Umemura}}
\affil{Center for Computational Physics, University of Tsukuba, \\
Tennoudai 1-1-1, Tsukuba, Ibaraki, 305-8577, JAPAN}
\email{umemura@rccp.tsukuba.ac.jp}
\and
\author{Hajime \textsc{Susa}}
\affil{Institute of Theoretical Physics, Rikkyo University, \\
Nishi-Ikebukuro 3-34-1, Toshima-ku, Tokyo, 171-8501, JAPAN}
\email{susa@rikkyo.ne.jp}

%


\maketitle

\begin{abstract}
Recently, ``CDM crisis'' is under discussion. 
The main point of this crisis is that number of substructures 
presented by cosmological N-body simulations based on CDM scenario 
for structure formation is much larger than observed substructures. 
Therefore, it is crucial for this crisis to discriminate whether
expected number of CDM substructures really exist but non-luminous 
or do not exist.

In this paper, we present a new idea to detect 
such invisible substructures by utilizing a gravitational lensing.
Here, we consider quasars that are  
gravitationally lensed by a foreground galaxy. 
A substructure around the lensing galaxy may 
superposed on one of the lensed images of such quasars.
In this situation, additional image splitting should occur 
in the image behind the substructure, and further multiple images are created. 
This is ``quasar mesolensing''. 

We estimate separation and time delay 
between further multiple images due to quasar mesolensing.
The expected value is $1 \sim 30$ milli-arcsecond for the separation 
and future fine resolution imaging enable us to find invisible substructures, 
and is $1 \sim 10^3$ second for the time delay  
and high-speed monitoring of such quasar will be able to find 
``echo''-like variation due to quasar mesolensing in intrinsic variability
of the quasar. 
Furthermore, we evaluate that the optical depth for the quasar mesolensing 
is $\sim 0.1$.
Consequently, if we monitor a few multiple quasars, 
we can find ``echo''-like variation in one of the images 
after intrinsic flux variations of quasars. 
\end{abstract}

\section{Introduction}

Cold Dark Matter (CDM) scenario for structure formation 
have been widely accepted in our Universe, and  
WMAP results (\cite{spergel}) also strengthened this scenario.
In addition, numerical simulations for structure formation 
based on CDM scenario nicely reproduce 
observed, large scale structures such as cluster of galaxies.  
However, as recently mentioned by \citet{klypin} and \citet{moore}, 
the scenario meets crisis in small scale structures. 
In their high resolution, 
cosmological N-body simulation based on CDM scenario, 
there are too many ``subhalos (or substructures)'' around galactic scale 
objects compared with actually observed substructures around Milky Way. 
If CDM scenario for structure formation is correct, 
many substructures should be invisible or not detectable 
due to very low star formation efficiency by some feedback processes 
(e.g., \cite{nishi}; \cite{kitay} or \cite{susa})
at least in current observational instruments. 
However, how can we confirm or reject the existence of 
theoretically predicted, many number of substructures around galaxies ?
The best probe can be gravitational lensing, 
since not brightness or luminosity of objects but only 
mass or density profile of objects is essential to this phenomenon.
Thus, we may be able to probe invisible substructures around galaxies 
by utilizing gravitational lensing. 

Recently, \citet{chiba} extends arguments of \citet{mao} 
and presents a nice idea. 
He focused on magnification anomalies in multiple images 
of two gravitationally lensed quasars. 
Such anomalies cannot be explained by smooth single lens model for 
lens galaxies and he proposed that the possible explanation for  
such flux anomalies is the existence of substructures around lens galaxies.
Following this work, \citet{dalal} investigate satellite mass fraction 
for seven gravitationally lensed quasars via Monte Carlo simulations, 
and they find that the resultant satellite mass fraction, $\sim 10~\%$ 
agree with predictions of cosmological N-body simulation based on CDM scenario
(see also \cite{metmad}, \cite{metzhao}). 
Furthermore, \citet{metmad} and \citet{metcalf} mentioned 
that milliarcsecond scale bending of radio jets 
in gravitationally lensed quasar is due to the distortion by 
gravitational lens effect of substructures around the lens galaxy. 

These works seem to find a way to save CDM scenario for structure formation 
from its crisis, but there still remains some ambiguity. 
First of all, flux magnification due to gravitational lens effect 
does not directly reflect mass of the lens, and magnification anomalies 
may not be direct evidence for substructures around lens galaxies. 
For example, as is well known in gravitationally lensed quasar 
with quadruple image, Q2237+0305 (Huchra's Lens or Einstein Cross), 
gravitationally lensed quasars may suffer quasar microlensing 
by stellar mass objects in the lens galaxy (\cite{ostensen}, 
see also \cite{osten2}, \cite{jackson}, \cite{rodorigo}, \cite{oshima} 
for quasar microlensing in another system).
Thus, magnification anomalies due to quasar microlensing 
can be occurring in most of gravitationally lensed quasars
\footnote{Surface mass density on images of gravitationally lensed quasars is 
estimated to be order of critical surface mass density of the lens systems. 
Additionally, the surface mass density is 
not so much different in different system. 
Therefore, optical depth for quasar microlensing should be order of unity, 
if most of the mass consists from stellar objects.}. 
However, the time scales for such phenomena in most systems are 
quite long compared with Q2237+0305, say several years, 
due to long distance to the lens galaxy, and 
it is difficult to discriminate magnification anomalies due to 
substructures around galaxies and that due to quasar microlensing 
only a few photometric observation. 
Secondly, image distortion of radio jets seems to stronger evidence 
for substructures around galaxies, but it can still not be direct 
evidence for such structures. 
The reason is the structure of radio jets is generally complicated 
and it may not easy task to find distortion between radio jets 
in corresponding images.
Of course, different from magnification, 
image distortion can be an indicator for typical scale of gravitational lens, 
e.g., Einstein ring radius, and also for mass of the lens.
However, the proposed idea is based on weak lensing regime, 
and it is not clear that the scale of distortion directly reflects 
the scale of gravitational lens, and 
there still remains some ambiguity, too.

In this paper, we investigate new important aspects of 
gravitational lensing to obtain more direct evidence for substructure 
around galaxies than previously proposed ideas.
Here, we focus on strong lensing regime of 
gravitational lens effects by substructures.
In this regime, further multiple images in one of multiple images 
of gravitationally lensed quasars are expected. 
This interesting phenomenon has never been discussed before,  
and we estimate expected values for the image separation and the time delay 
between such further multiple images. 
Reflecting mass of the lens objects, 
the values should be smaller than those for macrolensing 
due to galactic scale lens, but larger than those for microlensing 
due to stellar scale lens, 
and this gravitational lens effect can be called ``quasar mesolensing''.
Different from magnification anomaly, image separation and time delay  
between further multiple images induced by quasar mesolensing directly 
reflect mass of the lens objects via typical lens size such as 
Einstein ring radius for point mass lens, and these signals can be stronger 
evidences for the existence of substructures than magnification anomaly. 
In next section, we show that gravitationally lensed quasars are 
the most suitable targets to probe substructures around galaxies
from simple argument.
In section 3, lens models and numerical method are investigated, 
and the results of our calculations are presented in section 4.
Expected values for actual observations are presented in section 5, 
and final section is devoted to discussions.

\section{Suitable Lens System to Detect CDM Substructures}

In previous works, the detectability of substructure around galaxies 
are only considered in the case of gravitationally lensed quasars. 
Here, we go back to the first step and 
briefly discuss about the most suitable system to probe
substructure around galaxies.
Luminosity or surface brightness of substructures 
should be quite faint and unknown, and we investigate 
the method to probe substructures as not the source but the lens.
In the following argument, 
we fix total number of substructures ($N_{\rm sub}$), 
mass of individual substructures ($M_{\rm sub}$), 
and the size of host galaxies of substructures ($R_{\rm host}$) 
in all galaxies, for simplicity. 
If we take into account more realistic situations, 
the following argument for our current purpose will not dramatically change.  

The most essential quantity to estimate the detectability of 
gravitational lens effect is the optical depth 
for gravitational lensing, $\Upsilon$. 
This value is identical to the surface mass density of lens objects 
in the unit of critical surface mass density 
or the coverage fraction by the typical size of gravitational lens on the sky. 
If we put the host galaxy at the angular diameter distance $D_{\rm ol}$, 
substructures will exist roughly in the region with   
$S_{\rm host} \sim \pi (R_{\rm host} / D_{\rm ol})^2~{\rm radian}^2$. 
On the other hand, we consider the lens objects  as the point mass 
and the typical lens size is evaluated by 
Einstein ring radius ($\theta_{\rm E}$) which is expressed as 
\begin{equation}
\theta_{\rm E} = \left( \frac{4GM_{\rm sub}}{c^2} 
                  \frac{D_{\rm ls}}{D_{\rm ol} D_{\rm os}} \right) ^{1/2},  
\label{eq:einstein}
\end{equation}
where $D_{\rm ls}$ and $D_{\rm os}$ is the angular diameter distance 
between the lens and the source and the observer to the source, respectively.
By using this typical lens size, the total coverage 
by substructure lenses on the sky is 
$S_{\rm lens} \simeq N_{\rm sub} \cdot \pi \theta_{\rm E}^2 = 
4 \pi N_{\rm sub}GM_{\rm sub}D_{\rm ls} / (c^2 D_{\rm ol} D_{\rm os})
~{\rm radian}^2$. 
Consequently, the optical depth ($\Upsilon$) is evaluated as follows,
\begin{equation}
\Upsilon = \frac{S_{\rm lens}}{S_{\rm host}} \sim 
 N_{\rm sub} \frac{4GM_{\rm sub}}{c^2 R_{\rm host}^2} 
  \frac{D_{\rm ls} D_{\rm ol}}{D_{\rm os}}.
\label{eq:tau}
\end{equation}

The equation means that the distant lens system 
has larger optical depth for gravitational lensing 
except some special case such as self-lensing, 
i.e., $D_{\rm ol} \sim D_{\rm os}$. 
For example, the Gpc-scale lens systems have 
three orders of magnitude larger optical depth than 
the Mpc-scale lens systems.
Therefore, distant galaxies are more preferable to our purpose 
rather than the Milky Way galaxy, and gravitationally lensed quasars 
seem to be better candidate to probe substructures around galaxies.

Additionally, to utilize gravitational lensing, 
substructures as lens objects should be close to 
the line of sight to the sources. 
substructures are not distributed uniformly in universe 
but clustered around their host galaxy, 
and the line of sight to the source should intersect  
in the vicinity of galaxies.  
e.g., light from the source passes inside $R_{\rm host}$ 
from the center of the host galaxy. 
In this point of view, we can also say that 
gravitationally lensed quasars are 
the most suitable system to probe substructures, 
because such systems automatically satisfy this condition. 
We will present more detailed discussion about optical depth 
for strong lensing by CDM substructures in section 5.2.

\section{Calculation Methods}

Expected phenomena for gravitational lensing depend on 
the situation, lens model, some observational constraints and so on. 
To evaluate expected values for image separation and time delay 
between further multiple images, 
we describe adopted calculation methods for our estimations in this section.
Recently reported cosmological parameters measured by WMAP (\cite{spergel}) 
is adopted here, i.e., $\Omega_{m}=0.3$, $\Omega_{\Lambda}=0.7$, 
and $H_{0}=70~{\rm km~s^{-1}~Mpc^{-1}}$.
We set $z_{\rm l}=1$ (the lens redshift), $z_{\rm s}=2$ (the source redshift), 
$M_{\rm sub} = 10^7 M_{\odot}$ (or a corresponding velocity dispersion
\footnote{From the spherical collapse model that 
the collapse redshift is $\sim 20$, the velocity dispersion ($\sigma$) 
for such object is expressed as 
$v_{\rm sub} \sim 11.0 (M/10^7M_{\odot})^{1/3}~{\rm km~s^{-1}}$ 
(\cite{padma}).})
for a representative case unless otherwise specified. 

\subsection{Assumptions and constraints for estimations}

\begin{figure}
\begin{center}
 \FigureFile(150mm,150mm){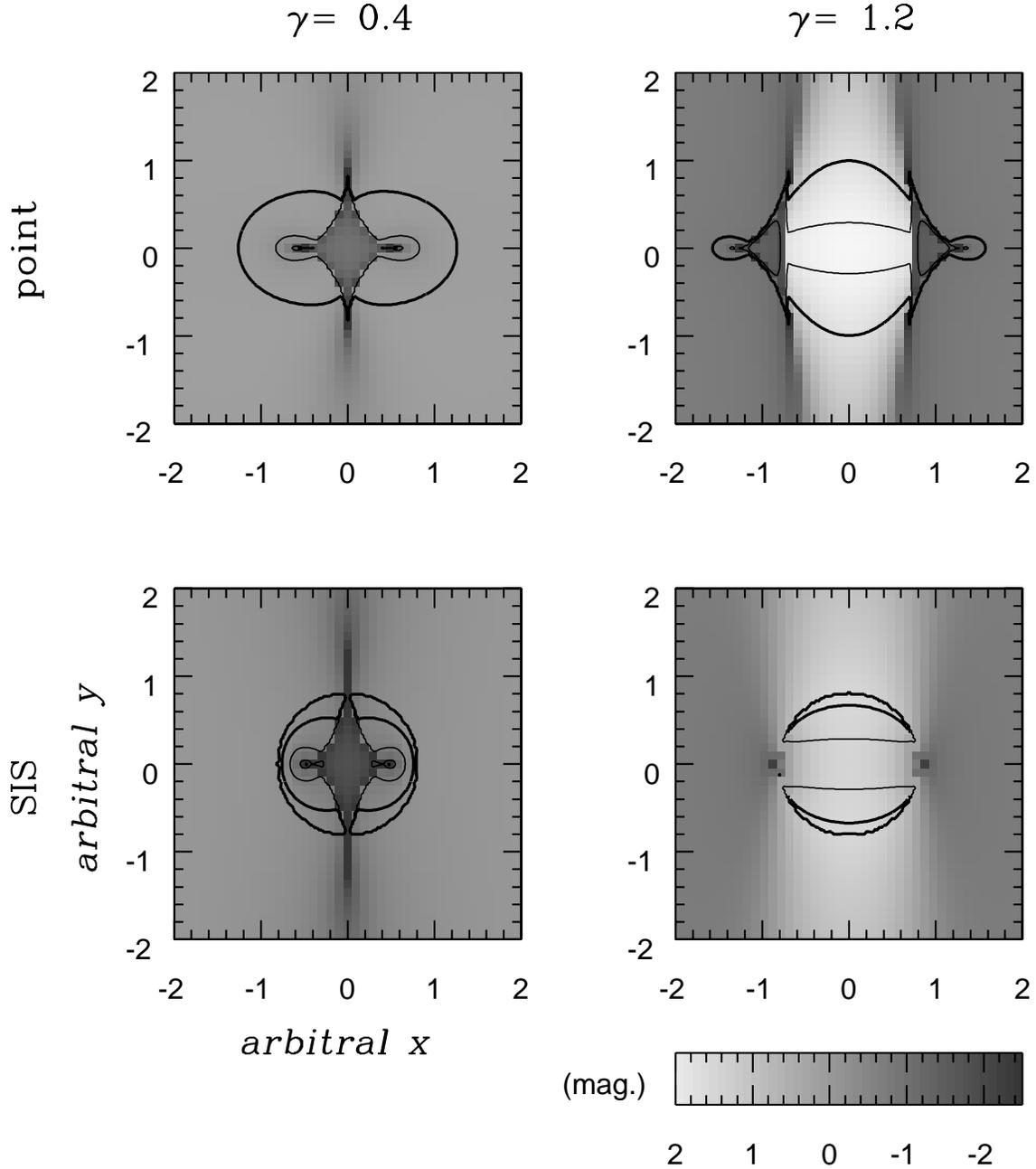}
\end{center}
\caption{Magnification patterns of ``Chang \& Refsdal'' lens 
for point mass lens (upper two panels) and for singular isothermal sphere 
(SIS) lens (lower two panels) are presented.
The gray scale is in the unit of magnitude as shown at the bottom.
Here, the external convergence ($\kappa$) is equal to zero and 
the external shear ($\gamma$) is $0.4$ in the left panels 
and $1.2$ in the right panels 
\footnote{These figures also valid for non-zero convergence case, and 
show magnification patterns for $\gamma_{\rm eff}=0.4$ and $1.2$. 
See section 4.2 for definition of $\gamma_{\rm eff}$,  
and appendix 2 for dependence on $\kappa$.}. 
The unit of $x-$ and $y-$axis is the typical lens size, e.g., 
Einstein ring radius for point mass lenses. 
Thin and thick lines show the contour for $f_{\rm lim}=2.0$ and $=10.0$, 
respectively, where $f_{\rm lim}$ is the flux ratio between 
the brightest two images (see text).}
\label{fig:contour}
\end{figure}

In the strong lensing regime for a single lens object, 
i.e., a substructure, the external effects may dramatically change 
the properties of the lens object as a gravitational lens. 
The most important effects are external convergence and shear. 
Generally, such external effects are caused by 
additional objects around the main lens object, a substructure in this case. 
There are two possible source to produce such effects in current situation. 
One is the host galaxy of substructures. 
As is well known in the argument for quasar microlensing and/or 
for constructing the macrolens model for gravitationally lensed quasars, 
the convergence around an image of gravitationally lensed quasars  
estimated to be order of unity (e.g., see \cite{schmidt}). 
Moreover, the shear is roughly the same value as the convergence. 
In such situations, even if the object is compact enough  
we can no more treat the main object as simple point mass lens, 
and we have to treat the lens as so-called ``Chang - Refsdal'' lens 
developed by \citet{charef}.
Particularly, the shear effect dramatically change the lens properties 
and magnification pattern becomes cuspy asteroid shape from circular shape, 
though the convergence effect is the simple focusing 
of the lens and the source planes.
Thus, we include external convergence and shear effect in our calculation. 
Since the typical lens size of substructures is much smaller than 
that of the host galaxy, 
we assume the external convergence and shear are constant. 
The magnification patterns in the case of ``Chang - Refsdal'' lens 
model are shown in figure~\ref{fig:contour}. 
Details for lens models are shown in section 3.2.

Another possible source to produce external effects is other structures,  
i.e., other substructures in the vicinity of the main substructure. 
If the lens plane is crowded with substructures and  
the optical depth is almost equal to or larger than unity, 
the gravitational lens properties such as magnification pattern will 
become very complicated similar to quasar microlensing (e.g., \cite{waps}). 
By using equation~\ref{eq:tau}, results obtained by numerical simulations 
(e.g., $N_{\rm sub} \sim 10^3$ inside $R_{\rm host} \sim 140~{\rm kpc}$), 
and put an arbitral mass for substructure, 
we are able to check the expected optical depth for substructure lens. 
There are some uncertainty due to lens properties, but 
if we set $M_{\rm sub} \sim 10^{7} M_{\odot}$, 
the expected value will become $\sim 10^{-4}$. 
As investigated by previous estimation (e.g., \cite{lees}), 
probability distribution for shear ($\gamma$) which produced by 
ensemble of point mass lenses with an average dimensionless 
surface mass density ($\kappa_{\ast}$), $\tilde{p}(\gamma, \kappa_{\ast})$, 
is expressed as 
\begin{equation}
 \tilde{p}(\gamma, \kappa_{\ast}) = 
  \frac{\kappa_{\ast} \gamma}{\left( \kappa_{\ast}^2 + \gamma^2 \right)^{3/2}}.
\end{equation}
Thus, cumulative probability distribution to produce a shear 
less than $\gamma$, $p(< \gamma | \kappa_{\ast})$, is evaluated as follows, 
\begin{equation}
 p(< \gamma | \kappa_{\ast}) = \int_{0}^{\gamma} 
  \tilde{p}(\gamma^{\prime}, \kappa_{\ast}) d\gamma^{\prime} 
   = 1 - \frac{\kappa_{\ast}}{\left( \kappa_{\ast}^2 + \gamma^2 \right)^{1/2}}.
\end{equation}
$\kappa_{\ast}$ is equivalent with $\Upsilon$ in current situation, 
and $p(< \gamma | \kappa_{\ast}) \sim 0.9$ for $\gamma \sim 10 \Upsilon$. 
In other words, effect of other substructures close to the main substructure 
is mostly less than $10^{-3}$ as an additional external shear. 
Even if we set $M_{\rm sub} \sim 10^{9} M_{\odot}$, 
the expected external shear will be $0.1$.
These values are small compared with external shear 
produced by the host galaxy of substructures, and 
we may safely be able to treat a substructure as an isolated lens object. 
More detailed arguments are presented in section 5.2, 
and will validate this ``single isolated lens'' treatment or 
``Chang - Refsdal'' lens treatment.   

As we have already noted in section 1, the target systems in our estimations, 
i.e., gravitationally lensed quasars, frequently suffer quasar microlensing 
and flux of images can be fluctuate around some median value
with fairly long time scale.
However, we are mainly focus not on magnification but 
on image separation and time delay between further multiple images. 
Since these two quantities are directly reflect the typical lens size 
and the size of stellar mass objects is several orders of magnitude 
smaller than that of substructures, 
the perturbation caused by stellar mass objects for quasar microlensing 
is negligibly small. 
Then, we do not include any effect due to quasar microlensing. 

\begin{figure}
\begin{center}
 \FigureFile(120mm,160mm){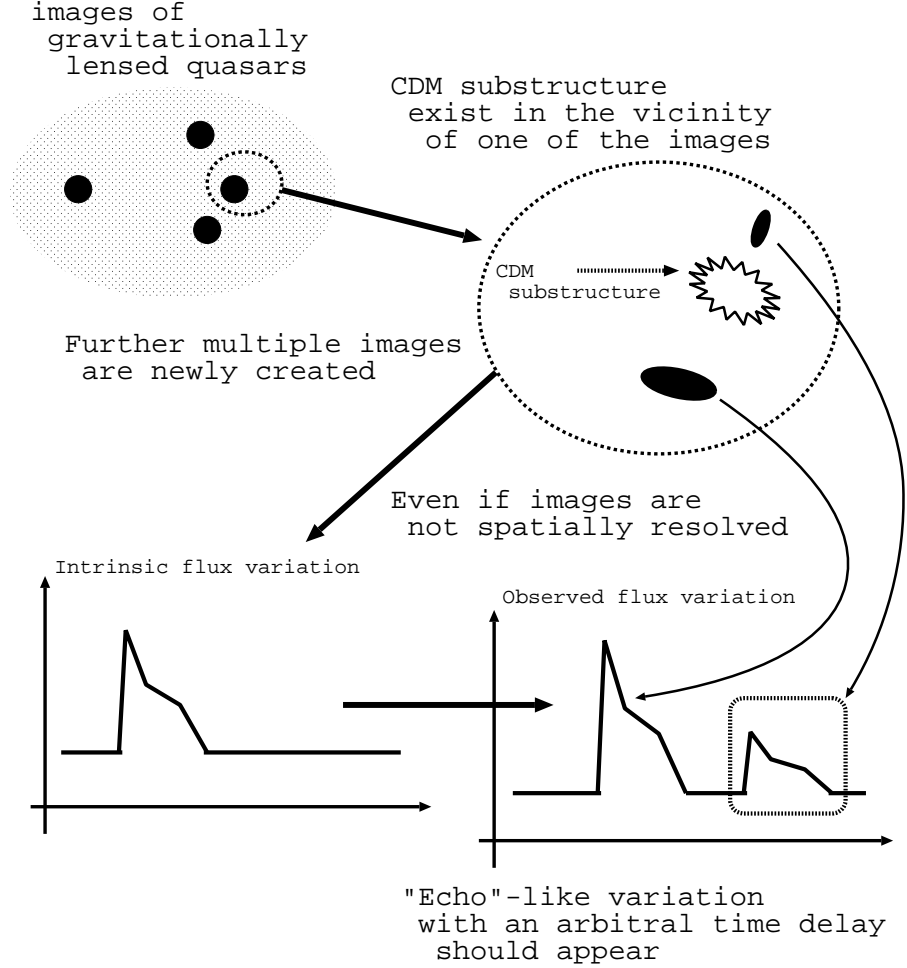}
\end{center}
\caption{Schematic picture of phenomena expected in strong lensing effects 
by substructures.}
\label{fig:schema}
\end{figure}

Finally, we have to mention about flux ratio 
between the primary brightest image ($A_1$) and 
the secondary brightest image ($A_2$), $f_{\rm lim} = A_1/A_2$, 
that is an important constraint to our estimations and 
will closely relate to the actual observational difficulties. 
To detect substructures by utilizing strong gravitational lensing, 
we have to hunt signals from further multiple images created by 
strong gravitational lens effect by CDM substructures. 
As schematically presented in figure~\ref{fig:schema}, 
we will find further multiple images beside one of multiple images 
of gravitationally lensed quasars with fine spatial resolution. 
Even if we cannot resolve further multiple images, 
we may be able to observe flux variation of a further multiple image 
superposed on that of another image. 
Such phenomenon will be observed as ``echo''-like flux variations 
in intrinsic variabilities of quasars. 
Since time lags of the ``echo''-like flux variation should be 
correspond to time delay between further multiple images, 
the measurement of the ``echo'' time lag is identical to the measurement of 
time delay between further multiple images.
Such measurement can be direct evidence for substructures around galaxies. 
However, actual observations suffer many kind of noises caused by 
detector, background objects, air glow and so on. 
Therefore, if the secondary brightest image is too faint compared 
with the primary brightest image, 
the ``echo''-like signal from the secondary brightest image 
may be smeared out in noises and 
we will not be able to find any evidence for substructure.  
To take into account such observational difficulties, 
we put some limits on $f_{\rm lim}$. 
Contours for some constant $f_{\rm lim}$ are presented in 
figure~\ref{fig:contour} for some cases.

\subsection{Lens models}

For ``Chang - Refsdal'' lens treatment, the lens equation 
with external convergence ($\kappa$) and shear ($\gamma$) is written as 
\begin{equation}
 \vec{\beta} = \left( 1 - \kappa \right) \vec{\theta} 
  - \gamma (\theta_x, -\theta_y) - \vec{\alpha}(\vec{\theta}),  
\label{eq:angle}
\end{equation}
where $\vec{\theta}=(\theta_x, \theta_y)$ and $\vec{\alpha}$ are 
arbitrary locations on the lens plane and bending angle at the location, 
respectively.
The shear direction in this formula is parallel to $x$-axis.
Solving this equation for a given $\vec{\beta}$, we obtain 
all the location of multiple images 
and can calculate separations between images.
For image 1 at location $\vec{\theta_1}$ and 
image 2 at location $\vec{\theta_2}$, 
the image separation ($\Delta \theta_{1,2}$) is calculated from 
$\Delta \theta_{1,2} = \left| \vec{\theta_1} - \vec{\theta_2} \right|$.  
By using lens potential ($\Psi$), arrival time delay ($\Delta t$)  
from the light path without gravitational lens effect is written as 
\begin{equation}
 \Delta t = \frac{1+z_{\rm l}}{2} \frac{1}{c} 
  \frac{D_{\rm ol}D_{\rm os}}{D_{\rm ls}} \left[ \left( \vec{\theta} 
   - \vec{\beta} \right)^2 - \kappa \left| \vec{\theta} \right|^2 
    - \gamma \left( \theta_x^2 - \theta_y^2 \right) - 2\Psi(\vec{\theta})
     \right].  
\label{eq:delay}
\end{equation}
Time delay between images is the arrival time delay difference 
between images, and we obtain time delay by subtracting 
$\Delta t$ at an image from $\Delta t$ at other images.
Thus, for image 1 at location $\vec{\theta_1}$ 
and image 2 at location $\vec{\theta_2}$, 
the time delay ($\tau_{1,2}$) is calculated from 
$\tau_{1,2} = \Delta t(\vec{\theta_1}) - \Delta t(\vec{\theta_2})$.     
Here, we apply two kind of lens models for substructures, 
one is point mass lens model and another is singular isothermal sphere (SIS) 
lens model (see section 6.1 for discussion about lens model).  

For point mass lens model, the typical lens size is given by 
Einstein ring radius, $\theta_{\rm E}$ (see equation~\ref{eq:einstein}), and  
the bending angle ($\vec{\alpha}$) at an arbitrary location 
on the lens plane ($\vec{\theta}$) is expressed as 
\begin{equation}
 \vec{\alpha} = \left( \frac{\theta_{\rm E}}{\left| \vec{\theta} \right|} 
  \right)^2 \vec{\theta} .
\end{equation} 
Additionally, the lens potential is given as follows 
\begin{equation}
 \Psi = \theta_{\rm E}^2 \ln \left| \vec{\theta} \right| . 
\end{equation}
For substructures with $M_{\rm sub} = 10^7 M_{\odot}$, 
$\theta_{\rm E}$ and $\theta_{\rm E}^2 
\frac{1+z_{\rm l}}{2} \frac{1}{c} \frac{D_{\rm ol}D_{\rm os}}{D_{\rm ls}}$ 
are equal to $4.23~{\rm mas}$ and $1.97 \times 10^2~{\rm s}$, respectively.

In contrast, for SIS lens model, 
the typical lens size, $\theta_{\rm SIS}$, is given as following form, 
\begin{equation}
 \theta_{\rm SIS} = 4 \pi \left( \frac{v_{\rm sub}}{c} \right)^2 
  \frac{D_{\rm ls}}{D_{\rm os}}, 
\end{equation}
where $v_{\rm sub}$ is the velocity dispersion of substructures. 
The bending angle and the lens potential in this case is presented as 
\begin{equation}
 \vec{\alpha} = \frac{\theta_{\rm SIS}}{\left| \vec{\theta} \right|}
  \vec{\theta}
\end{equation}
and  
\begin{equation}
 \Psi = \theta_{\rm SIS} \left| \vec{\theta} \right|,  
\end{equation}
respectively.
For substructure with $\sigma_{\rm sub} = 11.0~{\rm km~s^{-1}}$, 
$\theta_{\rm SIS}$ and $\theta_{\rm SIS}^2  
\frac{1+z_{\rm l}}{2} \frac{1}{c} \frac{D_{\rm ol}D_{\rm os}}{D_{\rm ls}}$ 
are equal to $1.27~{\rm mas}$ and $1.78 \times 10^1~{\rm s}$, respectively.

To evaluate actual values for given redshifts and mass 
or velocity dispersions, it is not efficient to perform calculations 
for all parameter combinations. 
As shown in appendix 1, image separation and time delay 
have scaling laws for typical lens sizes, 
and it must be convenient to utilize such scaling laws.

\begin{figure}
\begin{center}
 \FigureFile(170mm,60mm){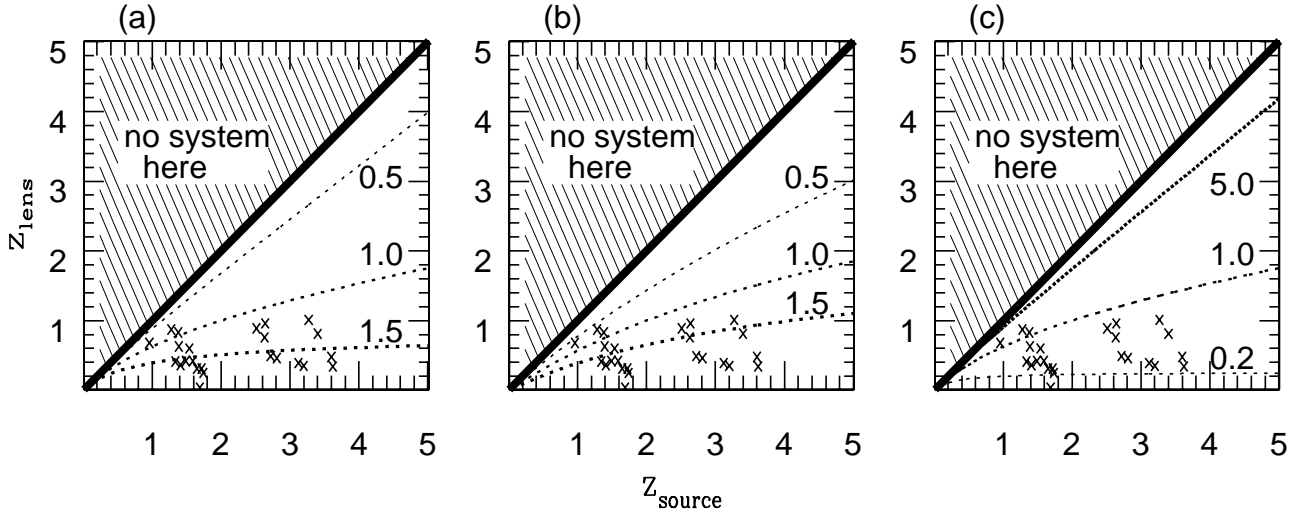}
\end{center}
\caption{The lens and the source redshifts dependences 
of typical lens size for point mass lens, 
$\theta_{\rm E}$ (left panel, a), for SIS lens (middle panel, b), 
and for the time delay (right panel, c) are presented. 
Dotted lines in left panel (a) and middle panel (b) 
show contours for $\theta_{\rm E}(z_{\rm l},z_{\rm s}) / 
\theta_{\rm E}(z_{\rm l}=1, z_{\rm s}=2)$ or 
$\theta_{\rm SIS}(z_{\rm l},z_{\rm s}) / 
\theta_{\rm SIS}(z_{\rm l}=1, z_{\rm s}=2)$
$=0.5$ (thin line), $1.0$ (middle line), and $1.5$ (thick line).  
Dotted lines in right panel (c) show 
contours for $(1+z_{\rm l})D_{\rm ol}D_{\rm os}/D_{\rm ls} = 0.2$ (thin line), 
$1.0$ (middle line), and $5.0$ (thin line) times 
larger than that for $(z_{\rm l},z_{\rm s})=(1, 2)$ for right panel (c). 
Cross symbols show currently known, redshifts measured lens systems. 
Since redshift of the lens should be smaller than that of the source, 
there is no lens system at the upper-left shaded region, 
$z_{\rm l} \ge z_{\rm s}$.}
\label{fig:zdeptot}
\end{figure}

Image separations are simply proportional 
to $\theta_{\rm E}$ for point mass lens and 
to $\theta_{\rm SIS}$ for SIS lens. 
For point mass lens, it is apparent that $\theta_{\rm E}$ is 
proportional to mass of substructures. 
System-to-system variance or redshifts dependence of $\theta_{\rm E}$ 
is shown in figure~\ref{fig:zdeptot} (a). 
At least for currently known lens systems, 
$\theta_{\rm E}$ values are factor of $\sim 2$ variance 
and system-to-system variance of expected image separations 
is not negligible but small. 
For SIS lens, $\theta_{\rm SIS}$ is proportional to square of 
velocity dispersion of substructures. 
System-to-system variance of $\theta_{\rm SIS}$ 
is also shown in figure~\ref{fig:zdeptot} (b).
Again, system-to-system variance is not negligible but small.  

Time delays are proportional 
to square of $\theta_{\rm E}$ for point mass lens, and 
to square of $\theta_{\rm SIS}$ for SIS lens.
Thus, dependence of $\theta_{\rm E}$ and $\theta_{\rm SIS}$ 
is $M_{\rm sub}^2$ for point mass lens and 
$\sigma_{\rm sub}^4$ for SIS lens, respectively. 
However, system-to-system variance is not so simple as 
the image separations, because we should take into account 
another term including system redshifts (e.g., see equation~\ref{eq:delay}). 
The effect of this additional term is presented 
in figure~\ref{fig:zdeptot} (c).
Gradient of this contour is roughly inverse of that 
in figure~\ref{fig:zdeptot} (a) and~\ref{fig:zdeptot} (b), 
and this additional effect will partly cancel out. 
Particularly, for point mass lens, angular diameter distances are 
completely cancel out and only $(1+z_{\rm l})$ term remains 
as system-to-system variance. 

Consequently, system-to-system variance is not so large, 
and our estimations present below will straightforwardly be applicable  
to any lens systems. 
If you require more detailed values, you will refer figure~\ref{fig:zdeptot}, 
and will simply multiply proper factors to our results.

\subsection{Numerical methods}

To estimate probability distributions for image separation 
and time delay between two brightest images, 
we performed Monte-Carlo simulation as follows;  

\begin{enumerate}

\item The source position is randomly determined within a circle. 
 The maximum radius for the source position is determined 
 that all region satisfies a required condition such as 
 constraint on $f_{\rm lim}$ is included. 

\item Solving the lens equation, we obtain all image positions 
 and corresponding magnification factors for the given source position. 
 Selecting the brightest two images, we calculate flux ($f$) ratio 
 between the brightest two images. 

\item If the flux ratio is smaller than a given constraint, 
 i.e., $f \le f_{\rm lim}$, the fainter image can also be observable and 
 we will continue further calculation for obtaining probability distributions.
 Therefore, resultant probabilities are conditional probabilities. 
 If the flux ratio is larger than a given constraint, 
 we will go back to the first step and perform the same calculation again 
 for other source position.

\item By using the source position which satisfy our constraint, 
 we can easily evaluate image separation, $\Delta \theta_{1,2}$, and 
 time delay, $\tau_{1,2}$, from equation~\ref{eq:angle} and 
 ~\ref{eq:delay}, respectively. 

\end{enumerate}

Repeat the above procedures (step $1 \sim 4$) for many times, 
and we obtain probability distributions for 
image separation and time delay. 
In this paper, we perform 50000 realizations. 
To achieve good accuracy for our calculations, 
the maximum radius for the source is large enough. 
On the other hand, to perform efficient calculations, 
the radius should not be too large.
Then, we determine the maximum radius for the source positions 
by referring magnification patterns and 
contours of flux ratio such as presented in figure~\ref{fig:contour}.
Though calculations for image separation and time delay 
are straight forward after we obtain image positions, 
calculations for image positions at a given source position are not, 
caused by the nature of lens equation. 
To overcome this difficulty, we adopt a proper method 
which utilizing so-called ``ray-shooting'' (e.g., \cite{keeton}).
Our numerical code solve the lens equation with sufficiently good accuracy.

\section{Individual Lens Properties}

Here, we present probability distributions for image separation and   
time delay for different $\gamma$, $\kappa$, and $f_{\rm lim}$ 
in the case of a single substructure, 
i.e., only one value for $\theta_{\rm E}$ and $\theta_{\rm SIS}$. 
Hereafter, cumulative distribution for quantity $x$ is presented 
as $p(<x|\kappa, \gamma)$. 
The reasons for dependences on the parameters 
are also discussed in this section. 
To make situation clear, 
we consider a few representative value for each parameter. 
Detailed values of our estimations or 
estimated values for other parameter sets are 
obtained via WWW. \footnote{ULR is 
{\tt http://www.rccp.tsukuba.ac.jp/Astro/yonehara/research.subhalo.html }}

\subsection{Effect of shear}

\begin{figure}
\begin{center}
 \FigureFile(80mm,100mm){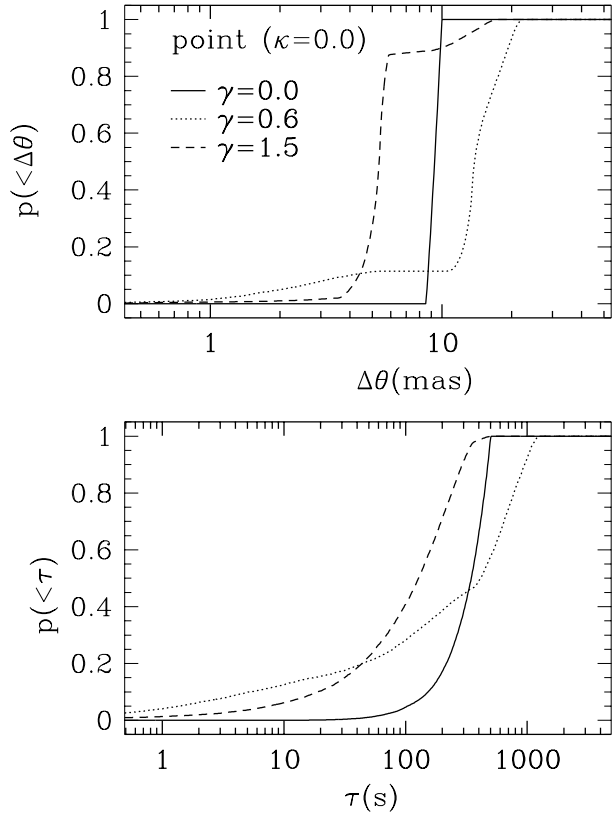}
\end{center}
\caption{Cumulative probability distribution of image separation 
(upper panel) and that of time delay (lower panel) are presented 
in the case of point mass lens model. 
Solid, dotted, and dashed lines indicate $\gamma=0.0$, $0.6$, and $1.5$, 
respectively.
Unit of the abscissa is ${\rm milli arcsecond}$ for image separation 
 and ${\rm second}$ for time delay. 
The scales of the abscissa are logarithmic.}
\label{fig:gdeppnt}
\end{figure}

First of all, we fix external convergence value as $\kappa=0$ 
and calculate probability distributions to investigate 
the effects of external shear for point mass lens case. 
Cumulative probability distribution of
image separation ($\Delta \theta$) and time delay ($\tau$) 
are presented in figure~\ref{fig:gdeppnt}.
As is clear from figure~\ref{fig:contour}, 
the lens properties dramatically change from $\gamma < 1$ case 
to $\gamma > 1$ case, and we present the result of 
$\gamma = 0.6$ as a representative case for $\gamma < 1$ 
and that of $\gamma = 1.5$ as a representative case for $\gamma > 1$.

Compared with no-shear case ($\gamma=0$), 
the shape of distribution dramatically changed 
and the range of distribution is extended toward 
large and small $\Delta\theta$ as depicted 
in upper panel of figure~\ref{fig:gdeppnt}. 
Though the lowest value of the distribution become small, 
the most part of the distribution shifted 
toward large $\Delta\theta$ direction by some factor. 
The shear effect will make the detection of quasar mesolensing signal easier. 
However, if the shear is too large and exceed unity, e.g., $\gamma=1.5$, 
the most part of the distribution shifted toward 
small $\Delta\theta$ direction by some factor. 
Therefore, too large shear will make the detection of 
quasar mesolensing signal more difficult. 

Since time delay is directly affected by image separation
(see equation~\ref{eq:delay}), the dependence on the external shear 
is similar to the image separation as depicted 
in lower panel of figure~\ref{fig:gdeppnt}. 
The shape of the distribution of time delay is apparently 
different from that of the image separation, 
but qualitative features are the same.
In large $\gamma$ case, the distribution extends  
toward large and small $\tau$ direction, 
and larger time delay will be expected compared with no-shear case.
On the other hand, the distribution shifted toward small $\tau$ direction 
in the case of $\gamma$ exceed unity again and the substructure detection 
via quasar mesolensing will become difficult.

\subsection{Model dependence}

\begin{figure}
\begin{center}
 \FigureFile(80mm,100mm){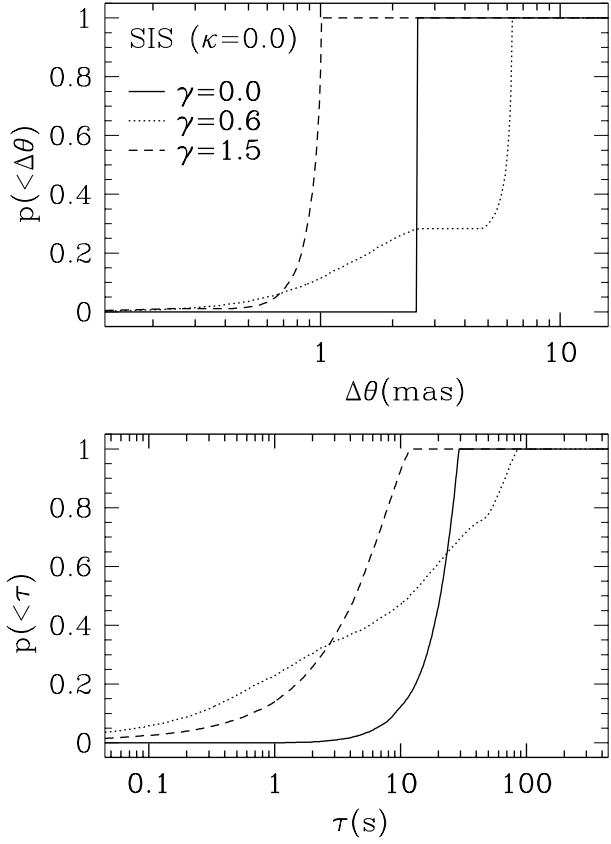}
\end{center}
\caption{Same as figure~\ref{fig:gdeppnt}, but for SIS lens model.}
\label{fig:gdepsis}
\end{figure}

Here, we present the expected values for SIS lens model. 
Before discussing the probability distribution, 
we briefly mention about basic properties of SIS lens model. 
It is well known that SIS lens produces pseudo-caustics 
\footnote{If the source crosses  pseudo-caustic 
(sometime it is called ``cut''), only one image will appear/disappear.
In contrast, if the source crosses usual caustic, 
two image with same flux but different parity will appear/disappear as a pair.}
at $|\vec{\beta}|=\theta_{\rm SIS}$. 
Generally, image separation increases with the source separate from the lens
(detailed explanations are presented in section 4.4). 
If the source locate outside both of caustics and pseudo-caustics, 
only one image appears, and we cannot define any quantities 
relate to our current estimations. 
Thus, different from the point mass lens, 
image pair cannot separate so large in the case of SIS, 
and the distribution shows sharp drop/rise at large value 
compared with point mass lens case, as presented in figure~\ref{fig:gdepsis}.

It is apparent that the resultant cumulative distributions show  
different shape from that for point mass lens case.
This is partly due to the unique nature of SIS lens model 
that we mentioned just before.   
Additionally, expected values are smaller than 
that for point mass lens case by a several factor. 
However, the dependence on external shear is qualitatively  
same as that for point mass lens case, 
i.e., large external shear makes expected values large but 
too large external shear, $\gamma \ge 1$, makes expected values small. 
There is no essential difference between these two lens models, 
and we mainly focus on point mass lens model at the following part.

\subsection{Effect of convergence}

As is already noted by \citet{falco}, external convergence leads 
degeneracy problem for modeling of lens objects. 
In contrast, if we fix lens model as current estimations, 
thanks to the simple nature of convergence, 
effects of external convergence can be easily take into account 
as proper scaling laws. 

Clearly, the convergence term in equation~\ref{eq:angle} will eliminate, 
if we divide both side of the equation by $(1-\kappa)$. 
After eliminating the convergence term, 
the lens equation becomes similar form again, 
and the shear term changes from 
$\gamma$ to $\gamma_{\rm eff} = \gamma / (1-\kappa)$.
Following some algebra as presented in appendix 2, 
we obtain the scaling laws of $\Delta \theta$, $\tau$, and 
$\mu_{\rm tot}$ (total magnification)
\footnote{Absolute values for all image are summed up.} 
for point mass lens model as  
\begin{eqnarray}
 p(<\Delta \theta|\kappa, \gamma) &=& 
  p \left( \left. <\frac{\Delta \theta}{(1-\kappa)^{0.5}} \right| 0, \gamma_{\rm eff} \right), \\
 p(<\tau|\kappa, \gamma) &=& p(<\tau|0, \gamma_{\rm eff}), \\ 
 p(<\mu_{\rm tot}|\kappa, \gamma) &=& 
  p \left( \left. <\frac{\mu_{\rm tot}}{(1-\kappa)^2} \right| 0, \gamma_{\rm eff} \right), 
\end{eqnarray}
and that for SIS lens model as 
\begin{eqnarray}
 p(<\Delta \theta|\kappa, \gamma) &=& 
  p \left( \left. <\frac{\Delta \theta}{(1-\kappa)} \right| 0, \gamma_{\rm eff} \right), \\
 p(<\tau|\kappa, \gamma) &=& 
  p \left( \left. <\frac{\tau}{(1-\kappa)} \right| 0, \gamma_{\rm eff} \right), \\ 
 p(<\mu_{\rm tot}|\kappa, \gamma) &=& 
  p \left( \left. <\frac{\mu_{\rm tot}}{(1-\kappa)^2} \right| 0, \gamma_{\rm eff} \right).
\end{eqnarray}
Therefore, to obtain the distributions for non-zero convergence and 
shear, $\gamma$, case, 
we calculate zero convergence with equivalent effective shear, 
$\gamma_{\rm eff}=\gamma / (1 - \kappa)$, case, 
and horizontally shift the resultant distributions by a proper factor.

\subsection{Flux ratio difference}

\begin{figure}
\begin{center}
 \FigureFile(80mm,100mm){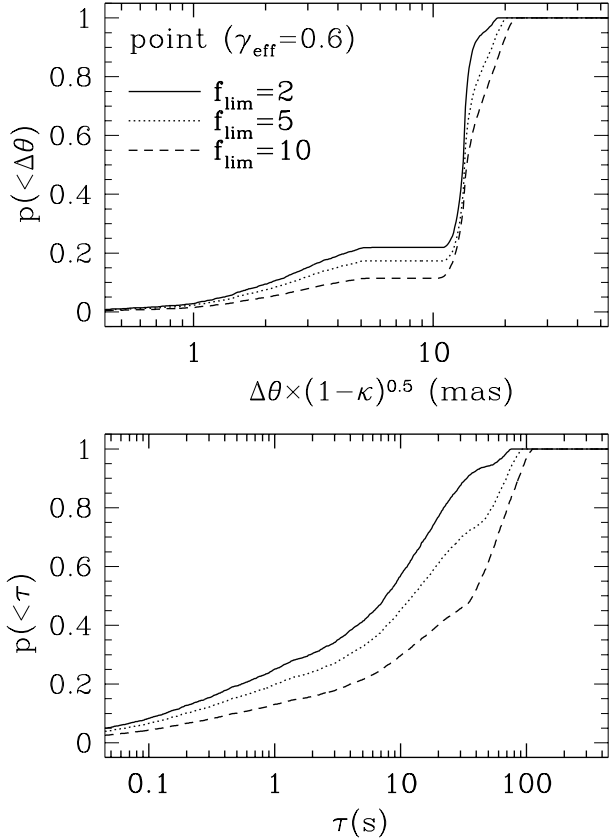}
\end{center}
\caption{Same as figure~\ref{fig:gdeppnt}, 
but for different $f_{\rm lim}$ case.
Solid, dotted, and dashed lines show $f_{\rm lim}=2$, $5$, 
and $10$ case, respectively.
The effective shear is fixed at $\gamma_{\rm eff} = 0.6$.} 
\label{fig:fdeppnt}
\end{figure}

At the end of this section, we present dependence on 
our assumed value for $f_{\rm lim}$ 
that will mimic constraint for real observations.
Here, we already have scaling laws for convergence, 
fixed shear value as $\gamma_{\rm eff}=0.6$ 
and present the resultant distributions for 
$f_{\rm lim}=2$, $5$, and $10$ case in figure~\ref{fig:fdeppnt}. 

In upper panel of figure~\ref{fig:fdeppnt}, image separation, 
the shape of the distribution does not change so much in horizontal direction 
but changes toward downward slightly with increasing $f_{\rm lim}$.
Then, the expected values becomes large with increasing $f_{\rm lim}$.
The reason is easily understood from the following analogy. 
The source separate enough from the lens, number of images becomes two. 
The more the source and the lens separates, 
the more these two images separates because 
one image is formed near the original source position, 
or the source position without lens effect, 
and another image is formed near the lens position. 
At the same time, since the image near the original source position 
becomes faint toward the original source flux, 
i.e., lower limit is the original source flux, 
since another image near the lens position 
becomes faint toward infinitely small flux, i.e., lower limit is zero. 
The image near the original source is usually brighter than another image, 
and finally, $f_{\rm lim}$ increase toward infinity with 
the source and the lens separation.
If we allow large value for $f_{\rm lim}$, 
it will be comparable to the situation that 
we can observe fainter images. 
In such a case, we can observe more separate image pairs 
as the above argument, and the expected image separation becomes large.
However, the fainter image becomes rapidly faint 
with the source and the lens separation, and change of the distribution 
is not significantly. 

Due to the same reason, the distribution of time delay 
only slightly changes with $f_{\rm lim}$ as clearly shown in 
lower panel of figure~\ref{fig:fdeppnt}.
Time delay relates to $\Delta\theta^2$ rather than $\Delta\theta$, 
and difference between the distribution for different $f_{\rm lim}$ 
seems to be large compared with that of image separation.
We note that the expected value becomes large 
with increasing $f_{\rm lim}$, again.

\section{Actual Lens Properties of CDM Substructures}

Employing basic properties for quasar mesolensing 
that presented in the previous section, 
we can evaluate really expected value for 
image separation and time delay due to quasar mesolensing
including statistical properties of substructures.

\subsection{Expected values}

To obtain expected values in a real situation, 
we have to include mass or velocity distribution of substructure. 
As shown previous sections, probability distributions are 
determined by $M_{\rm sub}$ (or $v_{\rm sub}$ for SIS lens) 
via the lens size, $\kappa$, $\gamma$, and $f_{\rm lim}$. 
If we put mass and velocity distribution 
\footnote{These are so-called ``mass function'' and ``velocity function'', 
and not cumulative ones.} 
for number of substructures as $n(M)$ and $n(v)$, respectively, 
really expected, cumulative distributions 
for point mass lens case $p_{\rm E}(<x)$ and 
for SIS lens case $p_{\rm SIS}(<x)$ will be written as
\begin{equation}
 p_{\rm E}(<x) = \frac{\int_{M_{\rm l}}^{M_{\rm u}} p_{\rm E}(M|<x) 
  n(M) dM}{\int_{M_{\rm l}}^{M_{\rm u}} n(M) dM}, 
\end{equation}
and
\begin{equation}
 p_{\rm SIS}(<x) = \frac{\int_{v_{\rm l}}^{v_{\rm u}} 
  p_{\rm SIS}(v|<x) n(v) dv}
   {\int_{v_{\rm l}}^{v_{\rm u}} n(v) dv}
\end{equation}
where, $p_{\rm E}(M|<x)$ and $p_{\rm SIS}(v|<x)$ 
are cumulative distributions for a given mass or velocity of 
substructures in the case of point mass lens and SIS lens, respectively.
$M_{\rm l}$, $M_{\rm u}$, $v_{\rm l}$, and $v_{\rm u}$ are 
lower limit and upper limit for mass of substructures and 
that for velocity of substructures, respectively. 

\begin{figure}
\begin{center}
 \FigureFile(80mm,100mm){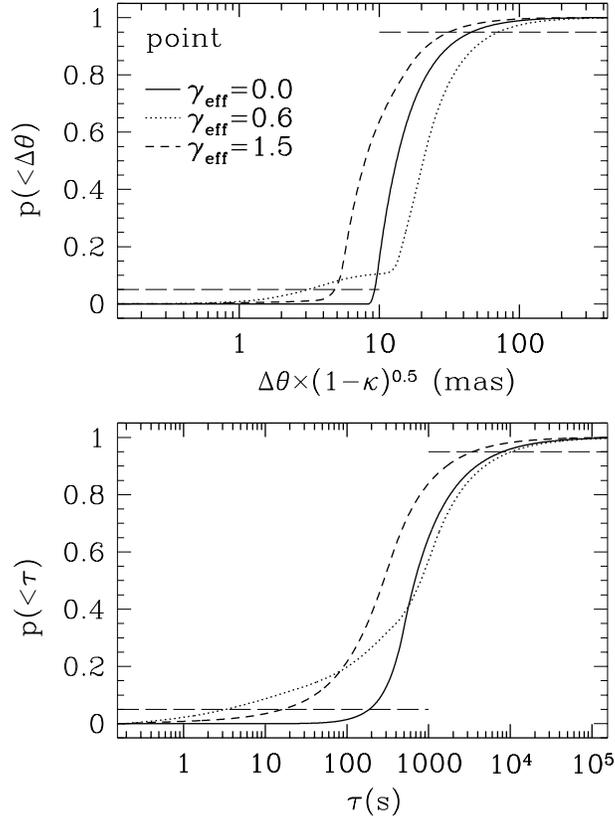}
\end{center}
\caption{Same as figure~\ref{fig:gdeppnt}, 
but for mass distribution for substructures is taken into account.
Thin long-dashed lines show $p(<\Delta\theta)=0.05$ (lower line) and 
$0.95$ (upper line). This is point mass lens case.} 
\label{fig:tdeppnt}
\end{figure}

\begin{figure}
\begin{center}
 \FigureFile(80mm,100mm){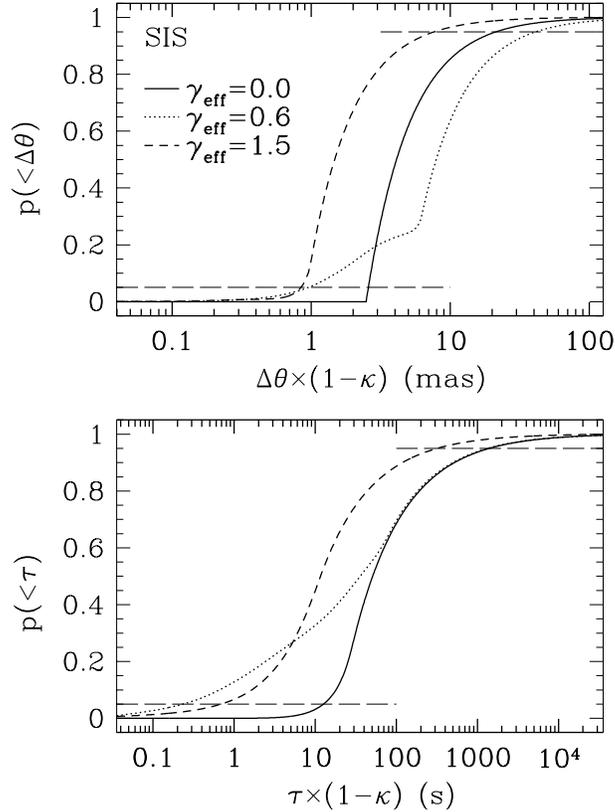}
\end{center}
\caption{Same as figure~\ref{fig:gdepsis}, 
but for velocity distribution for substructures is take into account.
Thin long-dashed lines show $p(<\tau)=0.05$ (lower line) and 
$0.95$ (upper line). This is SIS lens case.} 
\label{fig:tdepsis}
\end{figure}

Here, we assume power-law for $n(M)$ and $n(v)$, 
i.e., $n(M) \propto M^a$ and $n(v) \propto v^b$. 
Adopting the resultant power-law index obtained by cosmological 
N-body simulations (\cite{klypin}), we set $a = -1.92$ and $b = -3.75$. 
Upper limits for substructures are sufficiently above 
the resolution of the simulation. 
In contrast, lower limits are comparable to the current resolution 
(around $10^7M_{\odot}$ or $11~{\rm km~s^{-1}}$) and somewhat unclear. 
Of course, if we include small mass (or velocity) scale, 
number of substructures will increase and optical depth for 
quasar mesolensing will also increase. 
However, the expected image separation and time delay 
for such small scale structure should be small, 
and it can be difficult to detect. 
Therefore, we only take into account the mass and the velocity range 
for the previous simulations, and set $M_{\rm l}=10^7M_{\odot}$, 
$M_{\rm u}=10^{10}M_{\odot}$.
For the velocity range, we apply that the range corresponds to 
that of the mass range, and set 
$v_{\rm l}=11~{\rm km~s^{-1}}$ and 
$v_{\rm u}=110~{\rm km~s^{-1}}$.
The resultant distributions are presented in figure~\ref{fig:tdeppnt} 
for point mass lens case and in figure~\ref{fig:tdepsis}
for SIS lens case.
The distributions for various $\gamma_{\rm eff}$ 
are presented in these figures. 
In abscissa, we include $\kappa$ dependence discussed 
in section 4.3 for convenience.
For example, if we want to know the distribution of the image separation 
for $\kappa=0.5$ in the case of SIS lens model, 
$1~{\rm mas}$ and $1~{\rm s}$ in figure~\ref{fig:tdepsis} 
will correspond to $1~{\rm mas}/(1-0.5)=2~{\rm mas}$ 
and $1~{\rm s}/(1-0.5)=2~{\rm s}$, respectively. 

Comparing upper panels of figure~\ref{fig:tdeppnt} and~\ref{fig:tdepsis}, 
the expected image separations for point mass lens are slightly larger 
than that for SIS, but seems to be $\sim 1~{\rm mas}$ for lower limit 
and $\sim 30~{\rm mas}$ for upper limit. 
These values are comparable or smaller than 
spatial resolution of current observational instruments, 
but some future observational facilities will achieve this resolution
(see also discussion). 

The expected time delays are also shown in lower panels of 
figure~\ref{fig:tdeppnt} and~\ref{fig:tdepsis}. 
Reflecting the difference of image separation, 
the expected time delay for point mass lens is larger than 
that of SIS lens, again.
Difference between different lens model and different $\gamma_{\rm eff}$ 
is somewhat larger, but lower value for time delay is $\sim 1~{\rm s}$ 
and upper value is $\sim 10^3~{\rm s}$.
Of course, even in the case of $\sim 10^3~{\rm s}$, 
it is not easy task to find echo-like flux variation due to quasar mesolensing
in quasar variabilities with stochastic feature. 
However, this sub-day level time delay may be found by 
some coordinated monitoring observations in future (see also discussion).

\subsection{Probabilities for the strong lensing}

\begin{figure}
\begin{center}
 \FigureFile(100mm,140mm){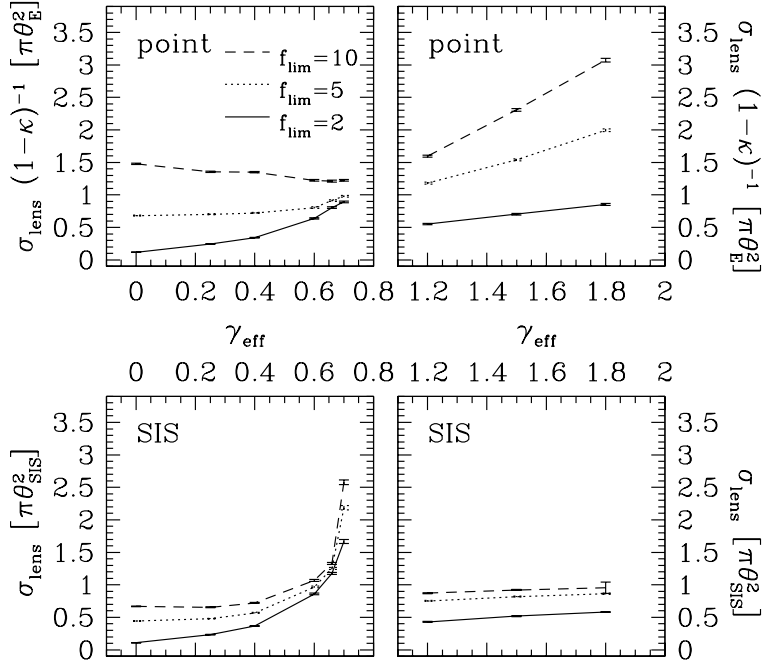}
\end{center}
\caption{Cross sections of quasar mesolensing 
for various $\gamma_{\rm eff}$ are presented in this figure.
Unit for the cross section is $\pi\theta_{\rm E}^2$ for point mass lens 
and $\pi\theta_{\rm SIS}^2$ for SIS lens.
Upper two panels show point mass lens case, and 
lower two panels show SIS lens case. 
Left panels are the case for $\gamma < 1$,  
though right panels are the case for $\gamma > 1$.
Solid, dotted, and dashed lines show the cross section for 
$f_{\rm lim}=2$, $5$, and $10$, respectively.} 
\label{fig:crssct}
\end{figure}

Reality for quasar mesolensing is evaluated by 
optical depth of gravitational lensing due to substructure.
It can be roughly estimated from equation~\ref{eq:tau}.
More exact optical depth are written as 
\begin{equation}
\Upsilon = \frac{1}{S_{\rm host}} \frac{1}{(1-\kappa)^2-\gamma^2}
 \int_{M_{\rm l}}^{M_{\rm u}} \sigma_{\rm lens}(M) n(M) dM, 
\label{eq:crsitgpnt}
\end{equation}
for point mass lens, and 
\begin{equation}
\Upsilon = \frac{1}{S_{\rm host}} \frac{1}{(1-\kappa)^2-\gamma^2}
 \int_{v_{\rm l}}^{v_{\rm u}} \sigma_{\rm lens}(v) n(v) dv, 
\label{eq:crsitgsis}
\end{equation}
where $\sigma_{\rm lens}$ denotes cross section of quasar mesolensing
for a given lens mass or lens velocity. 
The second fraction term in the right-hand side 
is the stretching effect of source plane 
and similar to so-called ``magnification bias''. 
Column number density of substructures is estimated in the lens plane,  
but the lens plane have already been magnified by a factor 
identical to this term, $\{ (1-\kappa)^2-\gamma^2 \}^{-1}$
\footnote{This factor corresponds to magnification factor 
due to the lens galaxy itself.}.
If we trace back to a region from the lens plane to the source plane, 
the surface area will become small by this factor.
Consequently, column number density of substructures  
is increased by a factor $\{ (1-\kappa)^2-\gamma^2 \}^{-1}$.
Cross section of quasar mesolensing 
for a given mass or velocity strongly depends on 
$\kappa$, $\gamma$, and other constraints ($f_{\rm lim}$ in this case).
Thus, it is difficult to calculate the cross section analytically 
in current situation (but see \cite{keeton2} for similar situations), 
and we evaluated it by using our Monte-Carlo simulation.
The results are shown in figure~\ref{fig:crssct}.

For $\kappa$ dependence, we can derive from scaling law for $\vec{\beta}$ 
between non-zero convergence case and zero convergence case 
with the same effective shear (see also appendix 2). 
In case of point mass lens model, $\vec{\beta}$ scaled by a factor of  
$(1-\kappa)^{0.5}$ from zero convergence case, and the cross section 
scaled by a factor of $\{ (1-\kappa)^{0.5} \}^2 = (1-\kappa)$.  
However, in case of SIS lens model,  
$\vec{\beta}$ does not alter by the effect of convergence, 
and the cross section is not affected by convergence. 
Then, $\sigma_{\rm lens}$ is proportional to $(1-\kappa)$ for point mass lens, 
and shows no dependence on $\kappa$ for SIS lens.
On the other hand, $\gamma$ dependence is complicated. 
Particularly, in $\gamma < 1$ regime, asteroid-shape caustics 
dramatically grows with the increase of $\gamma$. 
Of course, we can take into account such behavior 
into current estimation, but we simply set $\sigma_{\rm lens}$ equals to 
$\pi\theta_{\rm E}^2$ for point mass lens and 
$\pi\theta_{\rm SIS}^2$ for SIS lens. 
If the readers want to know more detailed values, 
figure~\ref{fig:crssct} can be the reference for this purpose.

From \citet{klypin}, $n(M) \simeq 19 (M/10^7M_{\odot})^{-1.92}$ 
and $n(v) \simeq 9.6 \times 10^3 (v/11~{\rm km~s^{-1}})^{-3.75}$
inside $140~{\rm kpc}$ radius 
\footnote{This correspond to $4.11~{\rm arcsec}$ in current situation.} 
of the lens galaxy. 
After we perform the integration in equation~\ref{eq:crsitgpnt} 
and~\ref{eq:crsitgsis}, we can obtain 
\begin{equation}
\Upsilon = 4.2 \times 10^{-2} \{ (1-\kappa)^2-\gamma^2 \}^{-1}
\end{equation}
for point mass lens and
\begin{equation} 
\Upsilon = 1.2 \times 10^{-2} \{ (1-\kappa)^2-\gamma^2 \}^{-1}
\end{equation}
for SIS lens.
In general, multiple images of gravitationally lensed quasars 
are magnified by a several or a few tenth factor 
that is identical to $\{ (1-\kappa)^2-\gamma^2 \}^{-1}$. 
Therefore, we can conclude that roughly ten percent of multiple images 
of gravitationally lensed quasars are affected by quasar mesolensing 
due to substructures.

The above values also depend on the spatial distribution of substructures.  
If substructures are concentrated within $1~{\rm arcsec}$ radius 
which is typical separation 
between the center of lens galaxy and multiple quasar images, 
column number density of substructures should be larger than 
that we used here, and the optical depth will increase 
from our results.


\section{Discussion}

\subsection{Comments on density profile of substructures}

Here, we mention about density profile of substructures.
To calculate gravitational lens effects and expected values, 
we require density profile for each substructure.
Unfortunately, even in the case of high resolution, cosmological 
N-body simulations for galaxy formation do not have enough 
mass resolutions for substructures, 
because an individual mass of a dark matter particle is 
too large and most of substructures are consists from 
a small number of dark matter particles (e.g., \cite{klypin}).
Thus, in this paper, we simply treat density profile of substructures 
as point mass or singular isothermal sphere. 
As is well known, properties of gravitational lens phenomena 
depend on density profile of lens objects, and 
differences of expected values between these two lens models  
reflect such dependence. 

In general, further multiple images are created when 
the density slope of lens object is smaller than $-1$.  
If NFW universal density profile (\cite{nfw}) holds 
even in the small objects such as substructures, 
size of substructures as lens object will be smaller than 
size of core radius of substructures. 
Consequently, the density slope of substructures is almost identical 
to that of core region of NFW-profile, $\sim -1$.
In this case, we cannot find any further multiple images 
caused by quasar mesolensing, and magnification anomaly can be 
a unique probe to detect substructures.
However, we still not have any evidence that 
NFW-profile holds in small scale such as substructures 
observationally and theoretically, and 
observational search for further multiple images caused by quasar mesolensing
must be a good approach to test density profile of substructures. 
If we find signals relate to further multiple images, 
that is observational evidence not only for existence of 
numerous substructures but also substructure has density profile 
which is different from NFW-profile. 
On the other hand, even if we fail to find such signals 
expect magnification anomaly, density profile of substructure 
should be similar to NFW-profile.

\subsection{Total magnification due to quasar mesolensing} 

\begin{figure}
\begin{center}
 \FigureFile(80mm,100mm){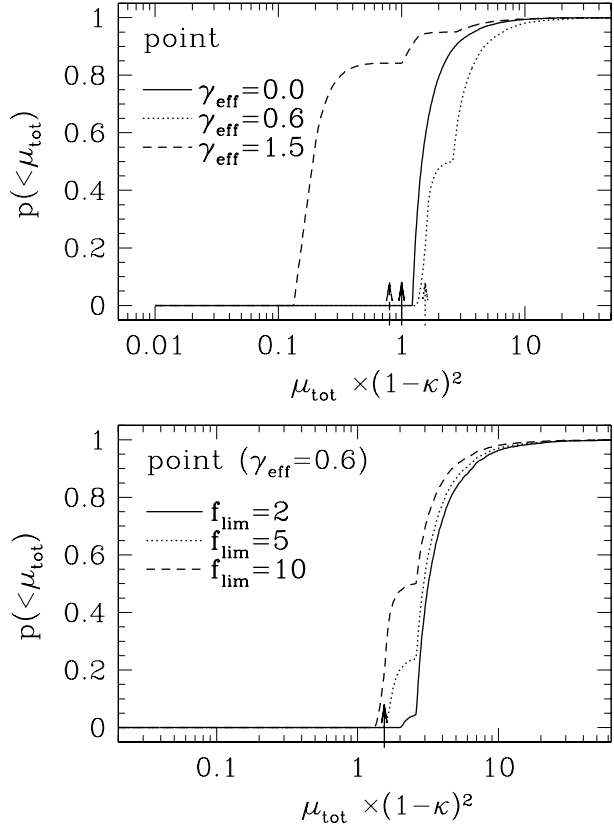}
\end{center}
\caption{Cumulative distributions of the total magnification 
for point mass lens case are presented in this figure. 
Upper panel shows shear dependence for $f_{\rm lim}=10$ case. 
Solid, dotted, and dashed line show $\gamma_{\rm eff}=0.0$, $0.6$, 
and $1.5$ case, respectively. 
Lower panel shows $f_{\rm lim}$ dependence for $\gamma_{\rm eff}=0.6$ case. 
Solid, dotted, and dashed line show $f_{\rm lim}=2$, $5$, 
and $10$ case, respectively.  
Unit of the abscissa is magnification factor and has no physical dimension.
Magnifications caused only by the external shear are
denoted by arrows with corresponding line type at the bottom of each panel.}
\label{fig:tmagpnt}
\end{figure}

\begin{figure}
\begin{center}
 \FigureFile(80mm,100mm){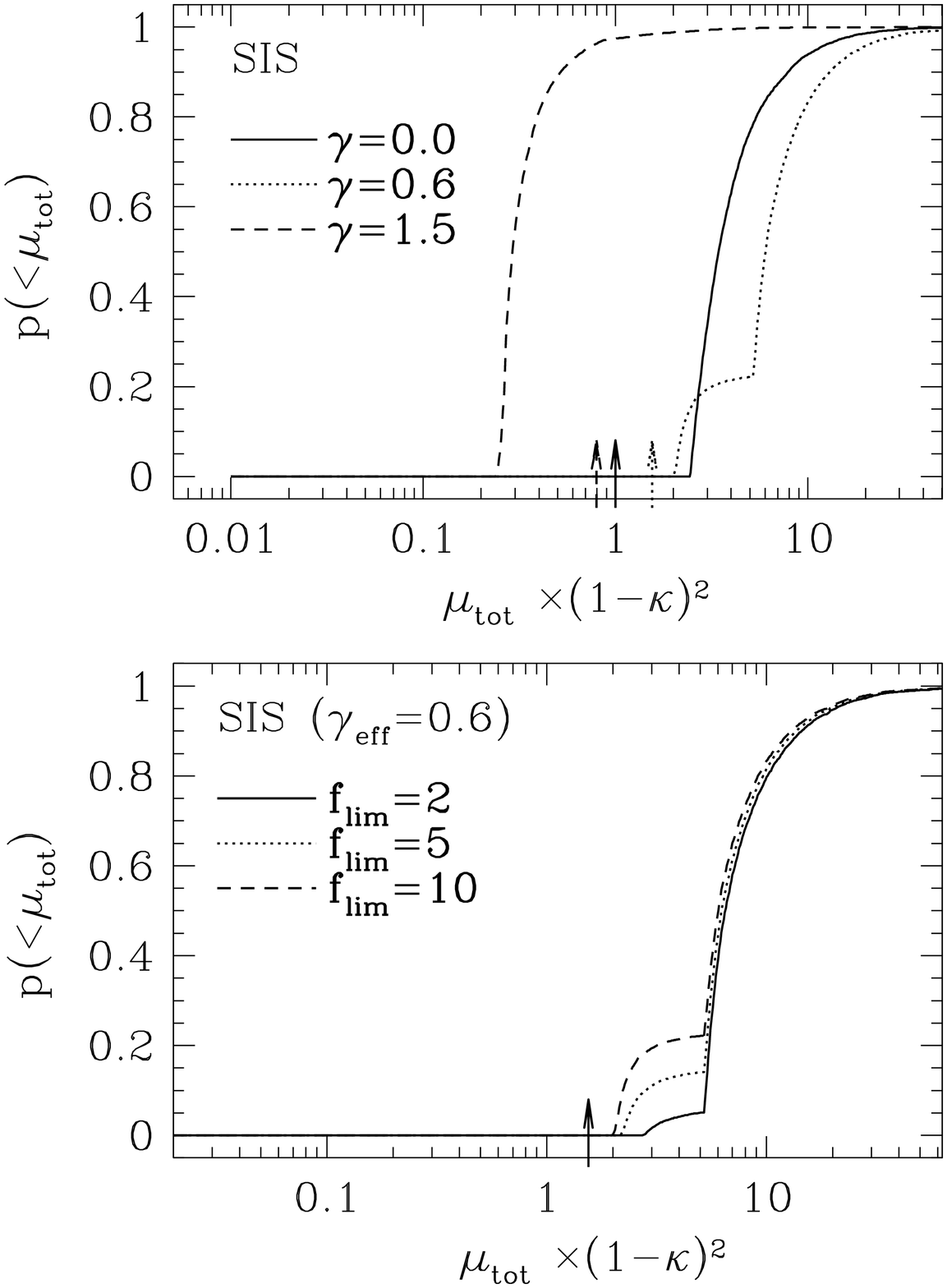}
\end{center}
\caption{Same as figure~\ref{fig:tmagpnt}, but for SIS lens case.}
\label{fig:tmagsis}
\end{figure}

The cumulative distribution of the total magnification 
depicted in figure~\ref{fig:tmagpnt} and~\ref{fig:tmagsis} for 
point mass lens case and SIS lens case, respectively. 
These figures are valid both of individual mass (or velocity) lens case 
and lens mass- (or velocity-) integrated case, 
because magnification does not include any information about 
lens mass (or velocity) and its properties are same 
for same convergence and shear case,  
even if we integrate over mass (or velocity) distribution of lens objects.  

Different from the distributions for the image separation and the time delay, 
these distributions consist from a few bump (or sharp rise/drop). 
This feature means that the expected value is clustered 
around a few location. 
When the external shear is added to axi-symmetric lens model,  
asteroid-shape caustics or highly magnified regions appear and 
such regions extend with increasing the shear value. 
Existence of caustics dramatically changes the magnification pattern, 
and the shape of distribution becomes to have two or more bump 
though only single bump appears in the case of no-shear. 
This feature is clearly seen in upper panels of 
figure~\ref{fig:tmagpnt} and~\ref{fig:tmagsis}. 

The bump at the high $\mu_{\rm tot}$ range corresponds to 
the distribution that the source is inside caustics. 
However, $\gamma_{\rm eff}$ exceeds unity,
caustics suddenly change from single asteroid shape to 
double triangle shape as shown in figure~\ref{fig:contour}. 
Following this dramatic change of magnification properties, 
demagnified region that never exist in the case 
of $\gamma_{\rm eff} < 1$ will appear, 
and the distribution widely extend toward $\mu_{\rm tot} < 1$ direction.
This feature can clearly be seen in $\gamma_{\rm eff}=1.5$ in upper panels 
of figure~\ref{fig:tmagpnt} and~\ref{fig:tmagsis}, again. 
In both lens model, expected value for the magnification 
reduced by a several factor. 
It has been already known that some combination of convergence and shear,  
particularly in the case of $\gamma_{\rm eff} > 1$, 
produces demagnified images. 
From our estimations, it becomes clear that 
the existence of substructures in lens galaxy sometime 
causes strong demagnification in an image of multiple quasars. 
If we find multiple quasars with odd number images  
\footnote{Usual multiple quasars have even number images}, 
such systems may be possible candidate for 
quasars with such strongly demagnified image, 
and this can be also useful to direct detection for 
quasar mesolensing and substructures.  

Here, we should note that these probability distributions for 
the total magnification show different properties from that 
presented by previous researches (e.g., \cite{keeton2}).
The reason is that the distribution in this paper include 
a condition to take into account observational constraint, $f_{\rm lim}$, 
though that in previous researches does not. 

For $f_{\rm lim}$ dependence of the total magnification,  
(lower panels of figure~\ref{fig:tmagpnt} and~\ref{fig:tmagsis}, 
the reason for the change is explained by the same argument 
as $f_{\rm lim}$ dependence of image separation in section 4.4.   
Large $f_{\rm lim}$ means that we can observe fainter image. 
The more the lens and the source separate, the larger $f_{\rm lim}$ becomes.
At the same time, the total magnification becomes small 
with increase of separation between the lens and the source, 
because flux of the brightest image approaches to the original source flux 
and flux of the secondly brightest image approaches to zero. 
Therefore, the distribution for large $f_{\rm lim}$ includes 
such source position case with low magnification (almost unity), and 
the probability at low $\mu_{\rm tot}$ increase.  
In contrast, the source position at high $\mu_{\rm tot}$ is close to the lens 
and flux of all image is comparable, i.e., $f_{\rm lim}$ is almost unity. 
Consequently, if we increase $f_{\rm lim}$, 
the source position with high magnification may not be newly included.
This is the reason why the shape at high $\mu_{\rm tot}$ shows 
no dramatic change. 
This result indicates that we may easily be able to detect quasar mesolensing 
signal in multiple quasar with weak magnification anomaly, 
if we achieve sufficiently high $f_{\rm lim}$ or can detect fainter images.

\subsection{Discriminate from quasar microlensing}

Further multiple images will be unambiguously detected,  
if the spatial resolution is sufficiently fine.
In contrast, for echo-like flux variation is not so simple. 
To confirm that the observed, echo-like flux variations 
are really due to quasar mesolensing, quasar microlensing seems to be 
somewhat confusing phenomenon. 
However, the time scale of echo-like flux variations 
is identical to the time scale of intrinsic quasar variabilities, 
and the shape of flux variations is also identical to that of intrinsic ones, 
because gravitational lens effect does not alter the time scale and the shape 
of flux variations, in principle. 
Only the difference between echo-like flux variations and 
intrinsic ones is the flux caused by a magnification factor difference. 
Moreover, the time scale of quasar microlensing 
is event time scales of microlensing event, 
and basically different from that of echo-like flux variations. 
Then, if we monitor all image of multiple quasars, 
we can easily obtain intrinsic flux variations of quasars 
at least from one of the images, because probability that 
all image suffers quasar mesolensing is less than one percent from 
previous estimates (section 5.2). 

Additionally, echo-like flux variations due to quasar mesolensing 
occur recurrently during a reasonable epoch, 
whereas quasar microlensing events occur occasionally. 
Time scale for quasar microlensing in usual system is 
expected to be a few year (\cite{waps}).
In contrast, duration of quasar mesolensing is roughly 
estimated to be $\sim \theta_{\rm E} D_{\rm ol} / V_{\rm sub}$, 
where $V_{\rm sub}$ is velocity of motion of substructures around galaxy. 
If we adopt $10^7M_{\odot}$ substructure and 
$200~{\rm km~s^{-1}}$ for $V_{\rm sub}$, the time scale becomes 
$ 5 \times 10^{12}~{\rm s} \sim 1.6 \times 10^5 ~{\rm yr}$. 
Duration of quasar mesolensing is sufficiently large 
compared with our life time, and we can treat quasar mesolensing 
as static gravitational lensing event. 
From this argument, echo-like flux variations due to quasar mesolensing 
is recurrent event, and the echo-like flux variations occur every time 
when the background quasars show their intrinsic flux variations.
This property makes us easier to detect echo-like flux variation 
and substructures via quasar mesolensing.

\subsection{Toward actual observation}

Quasar mesolensing occurs in $\sim 10~\%$ of images in multiple quasars, 
and we will be able to hunt such further multiple images 
as a direct evidence for substructures 
by using observational missions/facilities 
with $1 \sim 30~{\rm mas}$ spatial resolution 
or $1 \sim 10^3~{\rm s}$ time resolution. 
These expected values are not identical in different systems and images, 
because external convergence and shear due to the lens galaxy are 
different from system to system and from image to image. 
Therefore, to estimate the expected values more accurately 
in an individual system, we have to obtain $\kappa$ and $\gamma$ 
of images via modeling of the lens galaxy. 
Subsequently, calculate probability distributions for the observables 
as we shown in this paper. 
By comparing such probability distribution and observational results, 
we may be able to discuss about existence and nature of substructures.

To find further multiple images caused by quasar mesolensing directly,  
we require observational facility with 
$\le 30~{\rm mas}$-level spatial resolution. 
There is also non-zero probability to realize further multiple images 
with the separation $\ge 30~{\rm mas}$. 
Unfortunately, flux ratio of such image pairs is large,  
and the fainter image can be too faint to be detected feasibly. 
Thus, we really require $\le 30~{\rm mas}$ spatial resolution 
to detect quasar mesolensing. 
If we observe images with spatial resolution further below this value, 
we will clearly detect that the image is consisted by two or more 
images, and this can be the evidence for the existence of substructures. 
Even if we perform observation with marginal spatial resolution, 
we will find one or more structure is attached around the image, 
and this can be also the evidence for substructures. 
Simple way to achieve high spatial resolution is to use 
interferometers, but it may be practically difficult. 
The reason is that the light from multiple quasars images 
pass inside the lens galaxy, and the light suffers the effect 
such as scintillation due to relatively dense inter stellar medium. 
Therefore, not interferometric but direct observation is necessary 
and forthcoming observational instruments such as XMAS (\cite{kita})
\footnote{This will achieve $3~{\rm mas}$ spatial resolution in X-ray.} 
will be required. 

Another solution to find signal due to further multiple images 
is to hunt echo-like flux variations with $\le 10^3~{\rm s}$ delay. 
In this case, high spatial resolution is not required any more, 
but high time resolution or high speed monitoring is required. 
Observations of quasars that  show rapid and large flare-like 
flux variations are preferable to search echo-like signal. 
Generally, quasars always show stochastic flux variations,  
and it is practically difficult to find out echo-like flux variations 
due to quasar mesolensing of small flares with long duration.  
In contrast, large flares with short duration can be 
clearly detected from stochastic flux variations of quasars,  
and we can easily pick up corresponding echo-like flux variations 
due to quasar mesolensing.
Considering this point, for example, observation of rapid X-ray flares  
with time scale of ${\rm ks}$ (e.g., \cite{chartas}) that 
have recently detected in multiple quasars can be one of the best target.  
Furthermore, owing to shortness of the time delay, 
echo-like flux variation may be found by single observation 
with X-ray satellite.  
Duration of single observation with X-ray satellite is 
usually a several tenth of ${\rm ks}$, and sufficiently 
longer than expected time delay estimated in this paper.  
Thus, if we can detect such rapid X-ray flare and 
a substructure fortunately located in the vicinity of the image , 
we will be able to prove the existence of substructures 
by only single observation.
In the situation we are considering here, target quasar has multiple image, 
and we can discriminate echo-like flux variation due to quasar mesolensing 
from stochastic intrinsic flux variations in quasars 
by comparing flux variations in other images. 
For this confirmation purpose, multiple quasars with less than a day 
time delay between images are the ideal targets, because 
corresponding flux variations of delayed image also be able to observed 
by a single observation and other confusing phenomena such as 
quasar microlensing can be easily rejected.  

A rapid X-ray flare detected in RX~J0911.4+0511 by \citet{chartas} 
corresponds to $\sim 15~{\rm counts~s^{-1}}$ in CHANDRA. 
To detect echo-like flux variation with $3-\sigma$ confidence level, 
more than $9$ photons for a fainter image are required within a time bin. 
This $3-\sigma$ detection limits is almost comparable to 
$f_{\rm lim} = 15 / 9 \sim 2$ constraint. 
Referring figure~\ref{fig:fdeppnt}, $f_{\rm lim} = 2$ constraint 
reduces expected time delay by a several factor from $f_{\rm lim}=10$ 
constraint that we have applied to calculate figure~\ref{fig:tdeppnt}.
Even in this case, probability for $10^3~{\rm s}$ delay is non-zero, 
and we will be able to detect echo-like flux variations 
by using CHANDRA capability. 
For actual observation, it is better to perform such observation 
to multiple quasars with magnification anomaly than blind search, 
because magnification anomaly occurs also in this quasar mesolensing 
and such multiple quasars may have systematically high probability to 
show echo-like flux variation caused by quasar mesolensing 
than usual multiple quasars.  
The above discussion is limited on a case with $\sim 10^3~{\rm s}$ time delay. 
If there is no X-ray flares with larger amplitude than that \citet{chartas} 
have detected, we will have to wait X-ray facility 
with large collecting area such as XEUS 
\footnote{{\tt http://astro.esa.int/SA-general/Projects/XEUS/}}
or choose waveband with large signal-to-noise ratio for flares 
to find the evidence of substructures clearly and strongly.  
Expected time delay with shorter than $\sim 10^3~{\rm s}$ 
has significant fraction in cumulative distribution of the time delay 
as apparently shown in figure~\ref{fig:tdeppnt} and~\ref{fig:tdepsis}, 
and observational facilities with large collecting area or 
waveband that realize high signal-to-noise ratio will open a window 
to detect shorter time delay events.   

Until now, phenomena that we are investigated here are not detected, 
but some coordinated future observations enable us to reveal 
the existence of substructures around galaxies 
if CDM scenario for structure formation is totally correct.
 
~

The author would like to thank N. Yoshida, M. Chiba, and P.L. Schechter 
for their valuable comments and suggestions.
This work was supported in part by the Japan Society for
the Promotion of Science (09514, 13740124).

\clearpage

\appendix

\section{Dependence on Typical Lens Size}

Here, we consider the case when the lens size is 
multiplied by a factor ${\cal G}$ and derive 
scaling law of image separation and time delay between images.

\subsection{Point mass lens}

We put ${\cal G} \vec{\theta^{\prime}}$ and ${\cal G} \vec{\beta^{\prime}}$ 
instead of $\vec{\theta}$ and $\vec{\beta}$, respectively, 
the lens equation becomes
\begin{equation}
 {\cal G} \vec{\beta^{\prime}} = \left( 1 - \kappa \right) {\cal G} 
  \vec{\theta^{\prime}} - \gamma ({\cal G} \theta_x^{\prime}, 
   -{\cal G} \theta_y^{\prime} ) - \left( \frac{{\cal G} \theta_{\rm E}}
   {\left| {\cal G} \theta^{\prime} \right|} \right)^2 {\cal G} 
    \vec{\theta^{\prime}} .
\end{equation}
If we divide both side of the equation by ${\cal G}$, it reduce to 
\begin{equation}
 \vec{\beta^{\prime}} = \left( 1 - \kappa \right) \vec{\theta^{\prime}} 
  - \gamma (\theta_x^{\prime}, -\theta_y^{\prime} ) - 
   \left( \frac{\theta_{\rm E}}{\left| \theta^{\prime} \right|} \right)^2  
    \vec{\theta^{\prime}} ,
\end{equation}
and this equation is apparently the same form as the lens equation 
in the case of point mass lens for $\theta_{\rm E}$.
Thus, image separation between multiple images becomes 
$\vec{\theta_1} - \vec{\theta_2} = {\cal G} \left( \vec{\theta_1^{\prime}} 
 - \vec{\theta_2^{\prime}} \right)$ 
and is scaled simply by a factor of ${\cal G}$ 
compared with the case of $\theta_{\rm E}$.  

For arrival time delay, the equation becomes following, 
\begin{equation}
 \Delta t^{\prime} = \frac{1+z_{\rm l}}{2} \frac{1}{c} 
  \frac{D_{\rm ol}D_{\rm os}}{D_{\rm ls}} \left[ \left( 
   {\cal G} \vec{\theta^{\prime}} - {\cal G} \vec{\beta^{\prime}} \right)^2 
    - \kappa \left| {\cal G} \vec{\theta^{\prime}} \right|^2 
    \gamma \{ \left( {\cal G} \theta_x^{\prime} \right)^2 
     - \left( {\cal G} \theta_y^{\prime} \right)^2 \} 
      - 2 {\cal G}^2 \theta_{\rm E}^{\prime} 
       \ln \left| {\cal G} \vec{\theta^{\prime}} \right| \right].  
\end{equation}
All the term in the right hand side scaled by a factor of ${\cal G}^2$ 
except the last term. 
Fortunately, the observable is not arrival time delay but 
time delay between multiple images, $\tau_{1,2}$, 
and ${\cal G}$ inside $\ln$ in the last term vanishes 
when we subtract $\Delta t(\theta_2^{\prime})$ from 
$\Delta t(\theta_1^{\prime})$. 
Thus, the time delay between multiple images is scaled simply 
by a factor of ${\cal G}^2$ compared with the case of $\theta_{\rm E}$, 
i.e., $\tau_{1,2}^{\prime} = {\cal G}^2 \tau_{1,2}$.

The magnification factor for one image ($\mu$) is calculated from 
Jacobian matrix for $\vec{\beta} \rightarrow \vec{\theta}$ transformation, 
and written as 
\begin{equation}
 \mu = \left| \frac{\partial \vec{\beta}}{\partial \vec{\theta}} 
 \right|^{-1} = \left| \frac{{\cal G}}{{\cal G}} \frac{\partial  
  \vec{\beta^{\prime}}}{\partial \vec{\theta^{\prime}}} \right|^{-1} 
   = \left| \frac{\partial \vec{\beta^{\prime}}}
    {\partial \vec{\theta^{\prime}}} \right|^{-1}. 
\end{equation}
Thus, applying proper re-scaling for $\vec{\theta}$ and $\vec{\beta}$, 
there is no need to additional scaling for the magnification.

\subsection{SIS}

Applying the same scaling to $\vec{\beta}$ and $\vec{\theta}$  
as the point mass lens, the bending angle ($\vec{\alpha}$) and 
the lens potential ($\Psi$) for SIS lens model become 
\begin{eqnarray}
 \vec{\alpha}^{\prime} &=& \frac{{\cal G} \theta_{\rm SIS}}
  {\left| {\cal G} \vec{\theta^{\prime}} \right|} 
   {\cal G} \vec{\theta^{\prime}} \\ 
    &=& {\cal G} \frac{\theta_{\rm SIS}}{\left| \vec{\theta^{\prime}} \right|}
     \vec{\theta^{\prime}}
\end{eqnarray}
and 
\begin{eqnarray}
 \Psi^{\prime} &=& {\cal G} \theta_{\rm SIS}  
  \left| {\cal G} \vec{\theta^{\prime}} \right| \\
   &=& {\cal G}^2 \theta_{\rm SIS} \left| \vec{\theta^{\prime}} \right|,
\end{eqnarray}
respectively, and the same scaling law as point mass lens case 
holds also in this case. 
It is apparent that the scaling law for the magnification properties 
is same as the case for the point mass lens. 


\section{Dependence on Convergence}

Here, we derive scaling laws for convergence. 
We consider lens equations and arrival time delays for two cases, 
non-zero convergence ($\kappa$) and shear ($\gamma$) case and 
zero convergence and non-zero shear ($\gamma_{\rm eff}$) case. 
In the latter case, shear is equivalent with the effective shear 
for the former case, i.e., $\gamma_{\rm eff} = \gamma / (1 - \kappa)$.
We set the source position, the image position, arrival time delay, 
and the magnification factor for non-zero convergence case are 
$\beta$, $\theta$, $\Delta t$, and $\mu$, respectively, 
and that for zero convergence case are $\beta^{\prime}$, $\theta^{\prime}$, 
$\Delta t^{\prime}$, and $\mu^{\prime}$, respectively. 

\subsection{Point mass lens}

Here, we put $\theta_{\rm E}$ as the lens size for non-zero convergence case, 
and $\theta^{\prime}_{\rm E}$ as the lens size for zero convergence case. 
In this case, lens equations for both case are written as 
\begin{eqnarray}
 \vec{\beta} &=& (1-\kappa) \vec{\theta} - \gamma (\theta_x, -\theta_y) 
  - \left( \frac{\theta_{\rm E}}{|\vec{\theta}|} \right)^2 \vec{\theta}, 
 \nonumber \\
 &\rightarrow& \frac{\vec{\beta}}{(1 - \kappa)} = \vec{\theta} 
  - \gamma_{\rm eff} (\theta_x, -\theta_y) - \frac{1}{(1 - \kappa)}
 \left( \frac{\theta_{\rm E}}{|\vec{\theta}|} \right)^2 \vec{\theta}, 
  \label{eq:lenstrns1} \\
 \vec{\beta^{\prime}} &=& \vec{\theta^{\prime}} - \gamma_{\rm eff} 
  (\theta^{\prime}_x, -\theta^{\prime}_y) - 
   \left( \frac{\theta^{\prime}_{\rm E}}{|\vec{\theta^{\prime}}|} \right)^2 
    \vec{\theta^{\prime}}.
 \label{eq:lenstrns2}
\end{eqnarray}
If we set $\theta^{\prime}_{\rm E} = \theta_{\rm E} / (1 - \kappa)^{0.5}$
in equation~\ref{eq:lenstrns1}, the lens equation becomes similar form 
to equation~\ref{eq:lenstrns2}. 
From dependence on lens size presented in appendix 1.1, 
the source and the image positions of non-zero convergence 
case should be related to that of zero convergence case as follows, 
\begin{eqnarray}
 \frac{\vec{\beta}}{(1 - \kappa)} &=& 
  \frac{\vec{\beta^{\prime}}}{(1 - \kappa)^{0.5}} \label{eq:gdeptrns1} \\
 \vec{\theta} &=& \frac{\vec{\theta^{\prime}}}{(1 - \kappa)^{0.5}} .
\label{eq:gdeptrns2}
\end{eqnarray}
The above relation corresponding to scaling law for 
the source and the image positions between non-zero convergence case 
and zero convergence with the same effective shear case. 
Therefore, by using image separation for zero convergence case,  
$\Delta \theta^{\prime}_{1,2} = | \vec{\theta_1^{\prime}} - 
\vec{\theta_2^{\prime}} |$ , image separation for non-zero convergence case, 
$\Delta \theta_{1,2} = | \vec{\theta_1} - \vec{\theta_2} |$ is expressed as  
\begin{equation}
 \Delta \theta_{1,2} = 
  \frac{\Delta \theta^{\prime}_{1,2}}{(1 - \kappa)^{0.5}} .
\end{equation}

On the other hand, arrival time delays for both cases are written as 
\begin{eqnarray}
 \Delta t &=& \frac{1+z_{\rm l}}{2} \frac{1}{c} 
  \frac{D_{\rm ol}D_{\rm os}}{D_{\rm ls}} \left[ \left( \vec{\theta} - 
   \vec{\beta} \right)^2 - \kappa |\vec{\theta}|^2 - \gamma \left( 
    \theta_x^2 - \theta_y^2 \right) - 2 \theta_{\rm E}^2 \ln |\vec{\theta}| 
     \right] ,  \label{eq:atdtrns1} \\
 \Delta t^{\prime} &=& \frac{1+z_{\rm l}}{2} \frac{1}{c} 
  \frac{D_{\rm ol}D_{\rm os}}{D_{\rm ls}} \left[ \left( \vec{\theta^{\prime}} 
   - \vec{\beta^{\prime}} \right)^2 - \gamma_{\rm eff} \left( 
    \theta_x^{\prime~2} - \theta_y^{\prime~2} \right) - 
     2 \theta_{\rm E}^{\prime~2} \ln |\vec{\theta^{\prime}}| \right] .
\label{eq:atdtrns2}
\end{eqnarray}
By using equation~\ref{eq:gdeptrns1} and \ref{eq:gdeptrns2}, 
equation~\ref{eq:atdtrns1} becomes 
\begin{eqnarray}
 \Delta t &=& \frac{1+z_{\rm l}}{2} \frac{1}{c} 
  \frac{D_{\rm ol}D_{\rm os}}{D_{\rm ls}} \left[ \left( 
   \frac{\vec{\theta^{\prime}}}{(1 - \kappa)^{0.5}} - (1 - \kappa)^{0.5} 
    \vec{\beta^{\prime}} \right)^2 - 
     \kappa \frac{|\vec{\theta^{\prime}}|^2}{(1 - \kappa)} - \right. 
     \nonumber  \\
      &~& ~ \left. \frac{\gamma}{(1 - \kappa)} \left( \theta_x^{\prime~2} - 
       \theta_y^{\prime~2} \right) - 2 \theta_{\rm E}^2 
        \left( \ln |\vec{\theta^{\prime}}| - \ln |1-\kappa|^{0.5} \right) 
     \right] ,  \\
 &=& \frac{1+z_{\rm l}}{2} \frac{1}{c} \frac{D_{\rm ol}D_{\rm os}}{D_{\rm ls}} 
  \left[ \left( \vec{\theta^{\prime}} - \vec{\beta^{\prime}} \right)^2 - 
   \gamma_{\rm eff} \left( \theta_x^{\prime~2} - \theta_y^{\prime~2} \right) 
    - 2 \theta_{\rm E}^2 \ln |\vec{\theta^{\prime}}| \right. \nonumber \\
    &~& ~ \left. - \kappa |\vec{\beta^{\prime}}| + 
     \theta_{\rm E}^2 \ln |1-\kappa| \right].
 \label{eq:atdrslt}
\end{eqnarray}
Comparing with equation~\ref{eq:atdtrns2}, equation~\ref{eq:atdrslt} has 
4-th and 5-th terms in the parenthesis as extra terms. 
Fortunately, observable is not arrival time delay but 
arrival time delay difference between images, 
$\tau_{\rm 1,2} = \Delta t_1 - \Delta t_2$.   
These extra terms are constant and they cancel out 
in time delay between images. 
Therefore, time delay between images in the case of non-zero convergence, 
$\tau_{\rm 1,2}$, and that in the case of zero convergence with the same 
effective shear, $\tau^{\prime}_{\rm 1,2}$  are related as follows, 
\begin{equation}
 \tau_{\rm 1,2} = \tau^{\prime}_{\rm 1,2}.
\end{equation}

After this transformation, we can easily calculate scaling law for 
the magnification factor, $\mu$. 
As same as the lens size dependence, $\mu$ is expressed as 
\begin{equation}
 \mu = \left| \frac{\partial \vec{\beta}}{\partial \vec{\theta}} \right|^{-1} 
  = \left| \left( \frac{(1-\kappa)^{0.5}}{(1-\kappa)^{-0.5}} \right)^2 
   \frac{\partial \vec{\beta^{\prime}}}{\partial \vec{\theta^{\prime}}} 
    \right|^{-1} 
     =  \frac{1}{(1-\kappa)^2} \left| 
      \frac{\partial \vec{\beta^{\prime}}}{\partial \vec{\theta^{\prime}}} 
       \right|^{-1}
        =  \frac{1}{(1-\kappa)^2} \mu^{\prime}.   
\end{equation}
Thus, if we want to obtain the magnification factor
for non-zero convergence case, $\mu$, 
we will only multiply by a factor of $(1-\kappa)^{-2}$ to 
the magnification factor for zero convergence case, $\mu^{\prime}$. 
All image is multiplied by the same factor, $(1-\kappa)^{-2}$, 
and magnification ratio between two images 
does not alter in this transformation.

\subsection{SIS}

If we put $\theta_{\rm SIS}$ as the lens size for non-zero convergence case, 
$\theta^{\prime}_{\rm SIS}$ as the lens size for zero convergence case, 
and apply bending angle and lens potential for SIS lens model, 
we can lead scaling relation for convergence 
in the case of SIS lens model as same as the case of point mass lens. 

The form of bending angle and lens potential is different from 
the point mass lens case, $\kappa$ dependence of SIS lens case is 
different from the point mass lens case. 
Comparing lens equation for non-zero convergence case and 
that for zero convergence case, 
both lens equation will become similar form, 
if we set $\theta^{\prime}_{\rm SIS} = \theta_{\rm SIS} / (1 - \kappa)$. 
Consequently, scaling law for the source and the image position become 
followings, 
\begin{eqnarray}
 \frac{\vec{\beta}}{(1 - \kappa)} &=& 
  \frac{\vec{\beta^{\prime}}}{(1 - \kappa)}, \\
 \vec{\theta} &=& \frac{\vec{\theta^{\prime}}}{(1 - \kappa)} .
\end{eqnarray}
Therefore, relation between image separation for non-zero convergence case 
and that for zero convergence case is expressed as 
\begin{equation}
 \Delta \theta_{1,2} = \frac{\Delta \theta^{\prime}_{1,2}}{(1 - \kappa)} .
\end{equation}
Moreover, for arrival time delay, if we perform similar derivation 
to the case of point mass lens, 
relation between time delay between images for non-zero convergence case and 
that for zero convergence case is presented as 
\begin{equation}
 \tau_{\rm 1,2} = \frac{\tau^{\prime}_{\rm 1,2}}{(1 - \kappa)}.
\end{equation}

Again, scaling law for magnification factor is same as in the case 
of point mass lens, and magnification ratio between two images 
is identical when the effective shear is same.


%
%
%


\end{document}